\documentclass[preprint, 1p]{elsarticle}

\usepackage{caption}
\usepackage{graphicx}
\usepackage{subcaption}
\usepackage{placeins}
\usepackage{afterpage}
\usepackage{amsmath,amsthm,amssymb, microtype}
\usepackage[table,pdftex,dvipsnames]{xcolor}
\usepackage{float}
\usepackage{natbib}
\usepackage{booktabs, bm}
\usepackage[ruled,vlined]{algorithm2e}

\definecolor{bondiblue}{rgb}{0.0, 0.58, 0.71}
\definecolor{cerulean}{rgb}{0.0, 0.48, 0.65}

\usepackage[colorlinks]{hyperref}
\AtBeginDocument{%
  \hypersetup{
    citecolor=cerulean,
    linkcolor=cerulean,   
    urlcolor=cerulean}}

\usepackage[shortlabels]{enumitem}
\setlist[enumerate]{nosep}

\usepackage{lineno}

\usepackage{xargs} 
\newcommandx{\myparagraph}[1]{\paragraph{#1}}
\newcommandx{\mysubparagraph}{\bigskip \indent}

\journal{Neuroimage}
\setcitestyle{authoryear,open={},close={}}

\begin{document}
\begin{frontmatter}

\title{Multi-scale graph principal component analysis for connectomics}

\author[sw]{Steven Winter}
\ead{steven.winter@duke.edu}
\address[sw]{Department of Statistical Science, Duke University, Durham, NC, USA}

\author[zz]{Zhengwu Zhang}
\address[zz]{Department of Biostatistics and Computational Biology, University of Rochester, Rochester, NY, USA}

\author[dd]{David Dunson}
\address[dd]{Departments of Statistical Science and Mathematics, Duke University, Durham, NC, USA}

\begin{abstract}
In brain connectomics, the cortical surface is parcellated into different regions of interest (ROIs) prior to statistical analysis.  The brain connectome for each individual can then be represented as a graph, with the nodes corresponding to ROIs and edges to connections between ROIs.  Such a graph can be summarized as an adjacency matrix, with each cell containing the strength of connection between a pair of ROIs. These matrices are symmetric with the diagonal elements corresponding to self-connections typically excluded.  A major disadvantage of such representations of the connectome is their sensitivity to the chosen ROIs, including critically the number of ROIs and hence the {\em scale} of the graph.  As the scale becomes finer and more ROIs are used, graphs become increasingly sparse.  Clearly, the results of downstream statistical analyses can be highly dependent on the chosen parcellation.  To solve this problem, we propose a multi-scale graph factorization, which links together scale-specific factorizations through a common set of individual-specific scores.  These scores summarize an individual's brain structure combining information across measurement scales.  We obtain a simple and efficient algorithm for implementation, and illustrate substantial advantages over single scale approaches in simulations and analyses of the Human Connectome Project dataset.  
\end{abstract}

\end{frontmatter}


\section{Introduction}
With improvements in neuroimaging technology, there has been increasing interest in studying the brain connectome and how it varies across individuals, both randomly and in relation to individual-specific factors, ranging from genotype to neuro-behavioral traits [\cite{fornito2013graph}; \cite{van2013wu}; \cite{park2013structural}; \cite{miller2016multimodal}; \cite{glasser2016human}; \cite{tn-pca}].  In order to address such interests, it becomes crucial to obtain mathematical representations of the brain connectome that can be used routinely in statistical analyses.  By far the most common representation is via an adjacency matrix in which each cell of the matrix 
contains a measure of the strength of connection between a pair of regions of interest (ROIs).  In functional connectomics, the correlation in activity is often used as a measure of connection strength.  Although our methods are more widely applicable, our primary interest is in structural connectomics, which indirectly measures white matter (WM) fiber tracts connecting each pair of ROIs.  

A critical first step in analyzing structural connectomes is pre-processing of the raw imaging data to obtain a reconstruction of the white matter fiber tracts in each individual's brain, while also aligning these tracts into a common coordinate system [\cite{frost1}; \cite{Girard2014g}; \cite{tardif1}; \cite{glasser1}].  Then, before statistical analyses, one typically pre-specifies a particular set of ROIs; for example, corresponding to the Desikan-Killiany or Destrieux atlas [\cite{desikan-atlas}, \cite{destrieux-atlas}].  Based on these ROIs, a weighted or unweighted adjacency matrix can be defined for each individual under study; for example, with the cells of the matrix containing the number of fibers connecting each pair of ROIs in that individual.  Once these adjacency matrices are obtained, a rich variety of statistical methods are available.  These range from basing analyses on topological summaries of the network structure [\cite{bullmore1}; \cite{korgaonkar}], to comparing each connection separately possibly with multiple testing adjustment [\cite{zalesky}; \cite{meskaldji}], or more sophisticated methods that act on the entire adjacency matrices.  The latter class of methods include recently-developed approaches for replicated network analysis ranging from tensor PCA [\cite{tn-pca}] to Bayesian hierarchical models [\cite{airoldi}; \cite{rodriguez2012modeling}; \cite{miranda}; \cite{durante2017nonparametric}].  

A fundamental disadvantage of any approach based on defining ROIs is sensitivity to the chosen parcellation scheme. Sometimes it is partially based on physiological considerations of current knowledge of brain function, such as for the Desikan atlas, but typically the choice of parcellation scheme is largely arbitrary.  It is clear that statistical analyses can be very sensitive to the chosen parcellation: for a very fine scale parcellation of the brain, there are enormous numbers of ROIs so that the adjacency matrices become huge and sparse.  Analyses of such data may be fundamentally different than for more dense and coarse scale representations. A key insight is that different scale representations can be viewed as containing incomplete information about a single, common set of brain network latent factors.  The goal of this article is to develop simple and computationally efficient methods for routinely extracting such factors through a multi-scale graph PCA-type factorization, which takes as input brain connection graphs of a variety of scales. 

Our proposed approach builds upon \cite{tn-pca}, which proposed a tensor network factorization method for single-scale data.  Their method outputs a vector of real-valued brain structure latent factors for each individual under study; these factors can be used to visualize differences among individuals in their brain structure and in second stage analyses relating brain structure to other individual-specific variables.  Using their approach, they demonstrated a number of interesting relationships between brain structural connectivity and human traits and exposures in the Human Connectome Project (HCP) data.  However, this approach only works for single-scale network data and produces different results with different parcellations. To greatly reduce sensitivity to the network scale and definition of the ROIs, we propose a multi-scale generalization that can produce a common-set of individual-specific factors across scales.  This is shown to be fully flexible and to have substantial advantages in terms of measurement error-correction, combining information across scales, prediction of data on one scale from data on a different scale, and improved efficiency in analyses relating connectomes to other factors. 

Our proposed approach is conceptually related to ensemble learning methods, which combine multiple brain parcellations together into a single analysis [\cite{ensemble2}; \cite{ensemble3}; \cite{ensemble1}]. However, these approaches typically focus on combining different parcellations to obtain a black-box for prediction; our approach instead uses data from multiple different choices of ROIs to obtain a common set of brain principal components for each subject. These components can be used in highly interpretable analyses; for example, relating brain structure directly to traits as in \autoref{fig:mech}. We illustrate this through simulation experiments and applications to the Human Connectome Project dataset.  Code is freely available for public use at \url{https://github.com/szwinter/MultiGraph_PCA}.

\section{Data}

\subsection{The Human Connectome Project}

All data analyzed in this paper come from the Human Connectome Project (HCP), available via {\href{https://db.humanconnectome.org/}{ConnectomeDB}}. The HCP contains $1221$ high quality diffusion MRI (dMRI) scans from $1065$ unique healthy subjects - each scan is composed of $6$ runs summarizing three different gradient tables, with each table acquired once with right-to-left and then left-to-right phase encoding polarities. A typical gradient table contains $90$ diffusion weighting directions plus $6$ $b_0$ acquisitions distributed throughout each run. Each run is composed of three shells ($b=1000$, $2000$ and $3000 \text{ s}/\text{mm}^2$) with an equal number of acquisitions in each shell. All scans were performed with the same customized 3T Connectome Scanner. Basic processing yields final images with isotropic voxel size of $1.25$ $\text{mm}^3$. Please see \cite{glasser2} and \cite{van2013wu} for precise details.

\subsection{Connectome Extraction}
To facilitate summarizing structural connectivity (SC) utilizing different parcellations, we construct streamlines (white matter bundles) and extend them to the cortical surface. With parcellation labels on the surface, we can easily create the corresponding network adjacency matrix by summarizing the number of streamlines connecting a pair of ROIs. However,    
 most tractography algorithms cannot project streamlines to the surface due to two primary challenges. First, diffusion signals are unreliable in the gray matter (GM) and near GM-WM interface regions due to the anatomy of neurons, leading to an overestimation of streamlines in the WM that do not intersect the white surfaces [\cite{Girard2014g}; \cite{reveley2015superficialg}]. Second, there will be gyral biases in SC as shown in \cite{St-Onge2018g} due to a few factors, including low spatial resolutions during acquisition and biased tractography seeding choice. A recent development called surface-enhanced tractography (SET) [\cite{St-Onge2018g}] overcomes these challenges. {Instead of using the unreliable dMRI signal near the GM-WM interface region, SET uses a surface flow technique to model the superficial WM structure near the interface.} In SET, all reconstructed streamlines intersect with the white surface.

\begin{figure}[!t]
    \centering
    \includegraphics[width=\textwidth]{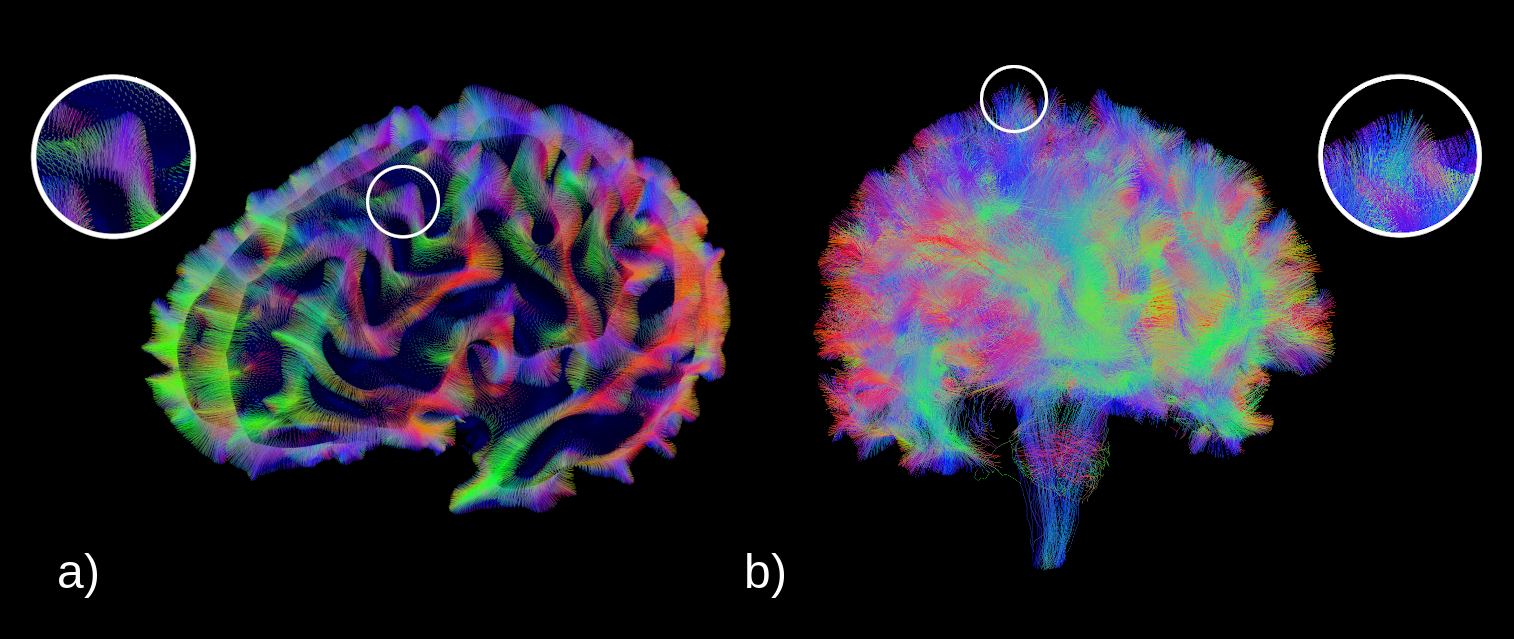}
    \caption{ Tractography results from SET. (a) shows the surface flow to one of the white surfaces, and (b) shows the final tractography with the reconstructed fanning structure near the white surfaces.}
    \label{fig:SET}
\end{figure}

In this study, we apply SET to build streamlines. The white surface is used to initiate the flow with a parameter $t$ controlling for the amount of flow into the WM, resulting in a surface beneath the white surface. Starting from the surface at $t > 0$, we initialize streamlines using $N_{s}$ seed points on the surfaces and propagate them using the particle filtering technique of \cite{Girard2014g}. We set $t=75$ and $N_{s} = 3\times10^6$ to obtain SC matrices according to the reproducibility results shown in \cite{cole2020surface}.  Figure \ref{fig:SET} illustrates some results from the SET pipeline, where (a) shows the initial white matter segments constructed by SET and (b) shows the final tractography result. Given the triangular mesh-based surfaces, a parcellation of the surfaces (e.g., Desikan atlas) and the tractography result, we summarize the streamlines between ROI pairs and obtain a relatively coarse adjacency matrix.    

\begin{figure}[t!]
\centering \includegraphics[width=\linewidth]{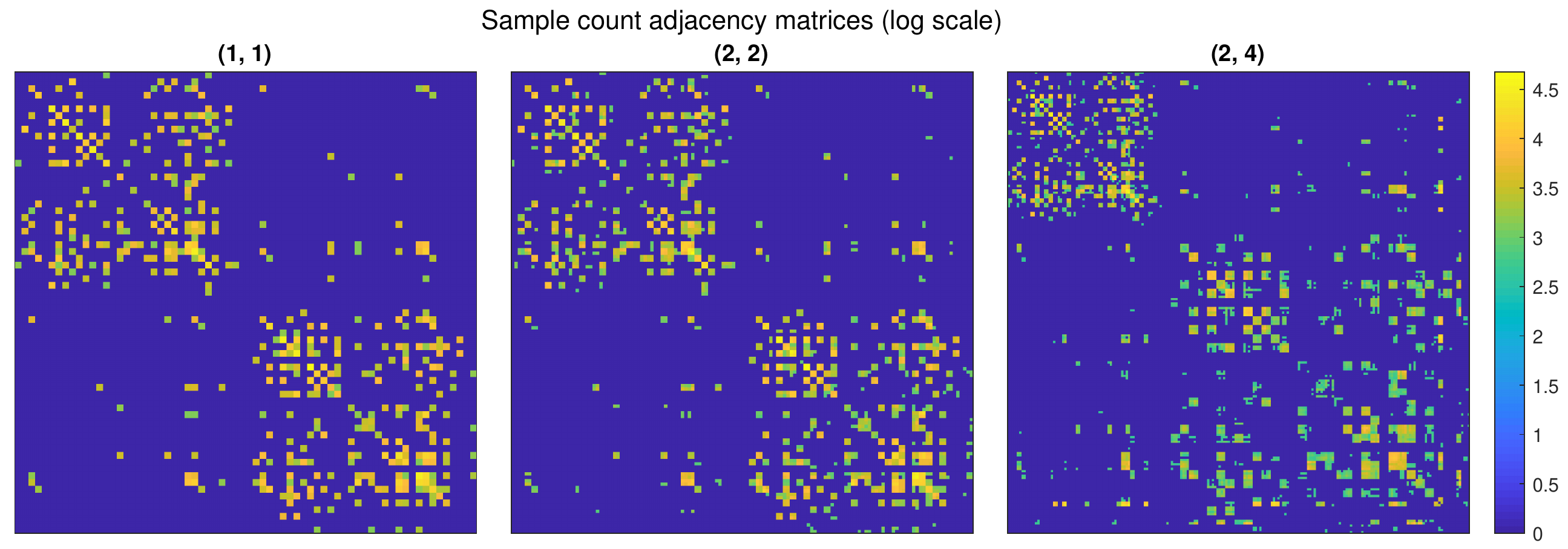}
\caption{Multiple adjacency matrix representations of a random individual's connectome. Notice the structural similarities between $(1, 1)$ and $(2, 2)$, contrasted with $(2, 4)$.}
\label{fig:examples}
\end{figure}

In practice, we find our multi-scale graph method performs best when the parcellations contain mostly different information instead of mostly redundant information. It is thus important to obtain a simple, reproducible method for generating very different parcellations - we propose to do this by dividing regions of the Desikan atlas into equally sized subregions. For simplicity we focus on custom parcellations parameterized by two natural numbers, $(\ell, r)$, indicating the number of new regions created from each Desikan region in the left and right hemispheres, respectively. Parcellations with $\ell=r$ will be referred to as $symmetric$ and parcellations with $\ell\neq r$ as $asymmetric$. For example, $(1, 1)$ corresponds exactly to the Deskian atlas; $(2, 2)$ splits every Desikan region into equally sized subregions; $(2, 4)$ splits every region in the left hemisphere into two equally sized subregions and every region in the right hemisphere into four equally sized subregions, resulting in a graph with $34\times 2 + 34\times 4 = 204$ nodes for the cortical surfaces. Examples of these graphs are visualized in \autoref{fig:examples}. 

Notice parcellations $(1,1)$ and $(2,2)$ show visually similar structures, hence they contain mostly redundant information about the connectome. Conversely, parcellations $(1,1)$ and $(2,4)$ look quite different - in particular, parcellation $(2,4)$ appears to be more sensitive to inter-hemisphere connections. In our experiments we find the greatest gains from multi-scale methods occur when both symmetric and asymmetric parcellations are used. Intuitively, the synthesis of these two disjoint views of the connectome allows us to inject more information about an individual's brain into the latent factors than, say, several symmetric parcellations.

\subsection{Trait Identification}
     The HCP contains rich trait measurement data for every subject in the study, quantifying cognitive, emotional, motor, sensory function and other domains. Most of the measurements were obtained in accordance with the NIH Toolbox for Assessment of Neurological and Behavioral function [\cite{gershon}]. Five additional traits were measured in HCP, including visual processing; personality and adaptive function; delay discounting; fluid intelligence; and behavioural measures of emotional processing. In this paper, we consider $175$ traits for each subject in our analysis, spanning eight categories: cognitive ability; motor skills; substance use; psychiatric and life function; senses; emotional ability; personality; and general health. Each trait is measured as a binary, ordinal, or continuous variable. A detailed description of these 175 traits can be found in \cite{behaviour}.

\section{Methods}

\subsection{Modelling}

Multi-scale graph data for individual $i$ comes as adjacency matrices $\{{\bf X}_i^{(1)},...,{\bf X}_i^{(R)}\}$, where $j=1,..,R$ indexes different parcellations of the cortical surface. Our goal is to replace these matrices with a real-valued vector of latent factors that summarizes the individual's brain, and then use this vector to study relationships between brain structure and traits. Of course there are many ways this can be done without multi-scale machinery: for example, one could vectorize the lower triangular part of each matrix \textit{within each scale} and use the resulting representations as inputs to linear or logistic regression with traits as responses. This would result in $R$ single-scale models which could be averaged or combined with e.g. Super Learner [\cite{superlearner}].
This approach can lead to successful prediction but suffers from a 
lack of interpretability. Specifically there is no principled method for visualizing ensemble-based inference on the level of connectomes, thus making such approaches sub-optimal for learning neural mechanisms underlying trait differences.

Our multi-scale model facilities interpretability by rephrasing the problem of computing latent factors as a problem of simultaneously approximating multiple graphs. Our particular $K$-dimensional approximation for individual $i$'s data takes the form
$$
    {\bf X}_i^{(j)} 
    = \sum_{h=1}^K  {\bf W}_h^{(j)} u_{i h}, \quad j=1,\ldots,R, 
$$    
where ${\bf W}_1^{(j)},\ldots,{\bf W}_K^{(j)}$ are independent rank $1$ adjacency matrices and ${\bf u}_i=(u_{i1},\ldots,u_{iK})$ is the row vector we use to summarize an individual's connectome. Notice the ${\bf W}$ matrices depend on the scale $j$ but not the individual $i$; conversely the latent factor vector ${\bf u}_i$ depends on the individual but not the scale. This dichotomy is critical: the graphs ${\bf W}_1^{(j)},\ldots,{\bf W}_K^{(j)}$ encapsulate scale-specific, population level graph structures - in essence, they provide the best $K$ dimensional approximation to an average connectome under parcellation $j$. The latent factors ${\bf u}_i$ encode \textit{scale-independent} individual deviations from the population average, hence providing a convenient summary of the unique features of an individual's connectome. Any inference performed with the latent factors can be mapped back on to the connectome level, allowing for clear, intuitive, and interpretable visualizations such as \autoref{fig:mech}.

\subsection{Implementation}

The multi-scale graph model presented above is over-parametrized. Each ${\bf W}_h^{(j)}$ is a rank $1$ adjacency matrix and can hence be decomposed as $d_h^{(j)} {\bf v}_h^{(j)}\circ {\bf v}_h^{(j)}$, where $d_h^{(j)}$ is a positive constant, ${\bf v}_h^{(j)}$ is a vector of unit length, and $\circ$ denotes the outer product.  We also impose within-scale orthogonality of the ${\bf v}_h^{(j)}$ to reliably estimate these separate contributors. With these constraints, our multi-scale graph model can be expressed as 
$$
    {\bf X}_i^{(j)} = \sum_{h=1}^K  d_h^{(j)} {\bf v}_h^{(j)}\circ {\bf v}_h^{(j)} u_{i h},\quad 
    j=1,\ldots,R, 
$$
which has considerably fewer free parameters compared with the original data. Finding optimal decompositions is nontrivial and necessitates expressing the problem in tensor notation. We adopt the conventions of \cite{kolda1}: henceforth scalars will be denoted by $x$; vectors by $\mathbf{x}$; matrices by $\mathbf{X}$. We will call $\mathcal{X}\in \mathbb{R}^{I_1\times I_2\times...\times I_M}$ an $M$-mode tensor; the inner product of two M-mode tensors $\mathcal{A}, \mathcal{B}\in  \mathbb{R}^{I_1\times I_2\times...\times I_M}$ is
\begin{align*}
    \langle \mathcal{A}, \mathcal{B}\rangle &= \sum_{i_1}\sum_{i_2}...\sum_{i_M}a_{i_1,i_2,...,i_M}b_{i_1,i_2,...,i_M} .
\end{align*}
This induces the $Frobenius$ norm via $||\mathcal{X}||_2 = \sqrt{\langle \mathcal{X}, \mathcal{X}\rangle}$. The \textit{n-mode multiplication} of a tensor $\mathcal{X}\in \mathbb{R}^{I_1\times I_2\times...\times I_M}$ with a matrix $A\in\mathbb{R}^{J_n\times I_n}$ (analogously, a vector) is denoted by $\mathcal{X}\times_n A$ and is defined as
\begin{align*}
    (\mathcal{X}\times_n A)_{i_1,...,i_{n-1},j,i_{n+1},...,i_M} &= \sum_{i_n}x_{i_1,i_2,...,i_M}a_{j, i_n},
\end{align*}
which is an element in $\mathbb{R}^{I_1\times...\times I_{n-1}\times J_n \times I_{n+1}\times...\times I_M}$.  

Concatenating the $N$ brain network matrices within each scale of data produces a semi-symmetric 3-mode tensor $\mathcal{X}^{(j)}\in \mathbb{R}^{P_j\times P_j\times N}$, where $P_j$ is the number of ROIs in the parcellation scheme $j$. For example, the complete collection of streamline count adjacency matrices from the HCP data under the Desikan parcellation would produce one such tensor of dimension $68\times 68\times 1065$, and splitting each Desikan ROI into two would produce another tensor of dimension $136 \times 136 \times 1065$. We can represent our multi-scale graph model at each scale as 
$$
    \mathcal{X}^{(j)}= \sum_{h=1}^K d_h^{(j)} \mathbf{v}_h^{(j)} \circ \mathbf{v}_h^{(j)} \circ \mathbf{u}_h,\quad j=1,\ldots,R.
$$
Constraining the $\mathbf{u}_h$ to be unit length allows one to solve each single scale approximation with a greedy, one-at-a-time rank-one algorithm referred to as the tensor power method [\cite{spca}]. The greedy algorithm results in the basis $\{d_h^{(j)}{\bf v}_h^{(j)} \circ {\bf v}_h^{(j)}, h=1,...,K\}$ that explains the most variation in the observed networks. 
Multilinear algebra is used to rephrase the greedy rank-one minimization problem as a maximization problem:
\begin{align*}
    \underset{\mathbf{v}_h, \mathbf{u}_h}{\text{maximize  }}&\mathcal{X} \times_1 \mathbf{P}_{h-1}\mathbf{v}_h \times_2 \mathbf{P}_{h-1}\mathbf{v}_h \times_3 \mathbf{u}_h \\
    \text{subject to } &||\mathbf{u}_h||_2 = 1, ||\mathbf{v}_h||_2 = 1
\end{align*}
where $\mathbf{P}_{h-1} = \mathbf{I} - \mathbf{V}_{h-1}\mathbf{V}_{h-1}^T$ with  $\mathbf{V}_{h-1}=[\mathbf{v}_1,\ldots,\mathbf{v}_{h-1}]$. Here the orthogonality constraint in the ${\bf v_h}$ manifests through the orthogonal projection matrices ${\bf \mathbf{P}_{h}}$. Obtaining a $joint$ approximation demands that we collapse these $R$ maximization problems into a single objective. We propose that one maximize the induced sum of squares. Specifically:
\begin{align*}
    \underset{\mathbf{u}_h, \mathbf{v}_h^{(1)},\ldots,\mathbf{v}_h^{(R)}}{\text{maximize  }}&\sum_{j=1}^R \bigg(\mathcal{X}^{(j)} \times_1 \mathbf{P}_{h-1}^{(j)}\mathbf{v}_h^{(j)} \times_2 \mathbf{P}_{h-1}\mathbf{v}_h^{(j)} \times_3 \mathbf{u}_h\bigg)^2 \\
    \text{subject to } &||\mathbf{u}_h||_2 = 1, ||\mathbf{v}_h^{(1)}||_2 = 1,\ldots,||\mathbf{v}_h^{(R)}||_2 = 1.
\end{align*}

To solve this maximization problem, blockwise coordinate ascent is sufficient: iteratively optimizing with respect to the $\mathbf{u}$ and $\mathbf{v}$ components yields closed form updates:
\begin{align*}
    \widehat{\mathbf{u}}_h|\mathbf{v}_h^{(1)},\ldots, \mathbf{v}_h^{(R)} &= E_{\max}\bigg(\sum_{j=1}^R\bigg(\mathcal{X}^{(j)} \times_1 \mathbf{P}_{h-1}^{(j)}\mathbf{v}_h^{(j)} \times_2 \mathbf{P}_{h-1}^{(j)}\mathbf{v}_h^{(j)}\bigg)\\
    &\hphantom{-------}\bigg(\mathcal{X}^{(j)} \times_1 \mathbf{P}_{h-1}^{(j)}\mathbf{v}_h^{(j)} \times_2 \mathbf{P}_{h-1}^{(j)}\mathbf{v}_h^{(j)}\bigg)^T\bigg) \\
    \widehat{\mathbf{v}}_h^{(j)}|\mathbf{u}_h &= E_{\max}\bigg(\mathbf{P}_{h-1}^{(j)}(\mathcal{X}^{(j)} \times_3 \mathbf{u}_h)\mathbf{P}_{h-1}^{(j)}\bigg),
\end{align*}
where $E_{max}(\mathbf{A})$ is the eigenvector corresponding to the maximum eigenvalue of the matrix $\mathbf{A}$. As the objective is not globally convex, there is the possibility of converging to a local optima. Warm starts can reduce this risk. In particular, we recommend computing the Higher Order Singular Value Decomposition (HOSVD) at each scale to obtain initial estimates of ${\bf v}^{(j)}$ and using these to find the first ${\bf u}$ update. We also advocate re-running the optimization routine several times for each rank $1$ problem; taking the best result across repeat runs further reduces the risk of selecting a local optima.

The form of $\widehat{\mathbf{u}}$ is appealing: using a sum of squares penalty results in an update rule that organically combines spectral information from all scales into a single vector. A linear penalty also yields a closed form update, but fails to incorporate any such spectral information. These developments are summarized in Algorithm 1.

\begin{center}
\begin{algorithm}[H]
\caption{Multi-scale graph principal components analysis.}
 \KwData{Tensors $\mathcal{X}^{(j)}\in \mathbb{R}^{P_j \times P_j \times N}$, $j=1,...,R$ and a positive integer $K$.}
 \KwResult{Vectors $\mathbf{d}^{(j)}=(d_1^{(j)},\ldots,d_K^{(j)})\in\mathbb{R}^K$ and matrices $\mathbf{V}^{(j)}=[\mathbf{v}_1^{(j)},\ldots,\mathbf{v}_K^{(j)}]\in\mathbb{R}^{P_j \times K}$, $\mathbf{U}=[\mathbf{u}_1,\ldots,\mathbf{u}_K]\in \mathbf{R}^{N\times K}$.} 
 \BlankLine
 \For{$j\leftarrow 1$ \KwTo $R$}{
$\widehat{\mathcal{X}}^{(j)}\leftarrow \mathcal{X}^{(j)}$
 }
 \BlankLine
  \For{$h\leftarrow 1$ \KwTo $K$}{
  initialize $\mathbf{u}_h$\;
  \For{$j\leftarrow 1$ \KwTo $R$}{
  initialize $\mathbf{v}_h^{(j)}$\;
  $\mathbf{P}_0^{(j)} \leftarrow {\bf I}_{P_j}$
 }
 \BlankLine
 \Repeat{objective converged}{
 $\widehat{\mathbf{u}}_h|\mathbf{v}_h^{(1)},\ldots, \mathbf{v}_h^{(R)} \leftarrow E_{\max}\bigg(\sum_{j=1}^R\bigg(\mathcal{X}^{(j)} \times_1 \mathbf{P}_{h-1}^{(j)}\mathbf{v}_h^{(j)} \times_2 \mathbf{P}_{h-1}^{(j)}\mathbf{v}_h^{(j)}\bigg)$\ $\hphantom{-------..-------} \bigg(\mathcal{X}^{(j)} \times_1 \mathbf{P}_{h-1}^{(j)}\mathbf{v}_h^{(j)} \times_2 \mathbf{P}_{h-1}^{(j)}\mathbf{v}_h^{(j)}\bigg)^T\bigg)$\;
  \For{$j\leftarrow 1$ \KwTo $R$}{
$\widehat{\mathbf{v}}_h^{(j)}|\mathbf{u}_h \leftarrow E_{\max}\bigg(\mathbf{P}_{h-1}^{(j)}(\mathcal{X}^{(j)} \times_3 \mathbf{u}_h)\mathbf{P}_{h-1}^{(j)}\bigg)$;
 }
 }
 \BlankLine
   \For{$j\leftarrow 1$ \KwTo $R$}{
  $\mathbf{V}_h^{(j)} \leftarrow [\mathbf{v}_1^{(j)},...,\mathbf{v}_h^{(j)}]$\;
  $\mathbf{P}_h^{(j)} \leftarrow \mathbf{I}-\mathbf{V}_h^{(j)}{\mathbf{V}_h^{(j)}}^T$\;
    $d_{h}^{(j)} \leftarrow \widehat{\mathcal{X}}^{(j)}\times_1\mathbf{v}_h^{(j)}\times_2\mathbf{v}_h^{(j)}\times_3\mathbf{u}_h$\;
  $\widehat{\mathcal{X}}^{(j)} \leftarrow \widehat{\mathcal{X}}^{(j)}-d_h^{(j)}\mathbf{v}_h^{(j)}\circ \mathbf{v}_h^{(j)} \circ \mathbf{u}_h$;
 }
}
\end{algorithm}
\end{center}

Interpretations of algorithm inputs/outputs and comments on connectome specific applications follow.

\subsection{Connectome Specific Interpretations}

Our multi-scale algorithm takes in tensors $\mathcal{X}^{(j)}\in \mathbb{R}^{P_j \times P_j \times N}$ for $j=1,\ldots,R$, each corresponding to a stack of $N$ single-scale brain network observations. It returns vectors $\mathbf{d}^{(j)}=(d_1^{(j)},\ldots,d_K^{(j)})\in\mathbb{R}^K$ and matrices $\mathbf{V}^{(j)}=[\mathbf{v}_1^{(j)},\ldots,\mathbf{v}_K^{(j)}]\in\mathbb{R}^{P_j \times K}$, $\mathbf{U}=[\mathbf{u}_1,\ldots,\mathbf{u}_K]\in \mathbf{R}^{N\times K}$. These are compactly stored as the sets $\{\mathbf{d}^{(j)}, \mathbf{V}^{(j)}, \mathbf{U}\},$ $j=1,\ldots,R$, which we informally refer to as $Kruskal$ decompositions. The rows of the latent factor matrix $\mathbf{U}$ summarize the connectomes of each individual - these are the outputs that can be used as features for statistical analysis. For the $j$th parcellation, the brain network modes $\{\mathbf{v}_1^{(j)}\circ\mathbf{v}_1^{(j)},\ldots,\mathbf{v}_K^{(j)}\circ\mathbf{v}_K^{(j)}\}$ explain more variation across subjects as $K$ increases. 

Our approach has a single tuning parameter $K$, which can be chosen using strategies for choosing the number of components in PCA. If the primary goal is to extract brain principal components predictive of traits of the individual, then choosing $K$ to maximize predictive accuracy out-of-sample is a reasonable strategy.  Alternatively, if the primary goal is to obtain a parsimonious representation of the brain connectome that discards limited information, then $K$ can be chosen so that at least some prespecified proportion of the variance is explained [\cite{spca}]. Formally, the multi-scale cumulative proportion of variance explained (CPVE) by the first $K$ components is given by
\begin{align*}
    \text{CPVE} = \underset{j}{\text{min  }} \frac{||{\bf\mathcal{X}}^{(j)}\times_1 {\bf P_{V_K^{(j)}}} \times_2 {\bf P_{U_K^{(j)}}}||}{||{\bf\mathcal{X}}^{(j)}||},
\end{align*}
where ${\bf P_{V_K^{(j)}}}$ (resp. ${\bf P_{U_K^{(j)}}}$) is the projection onto the first $K$ columns of ${\bf V}^{(j)}$ (resp. ${\bf U}^{(j)}$).

\begin{figure}[t!]
\centering \includegraphics[width=0.8\linewidth]{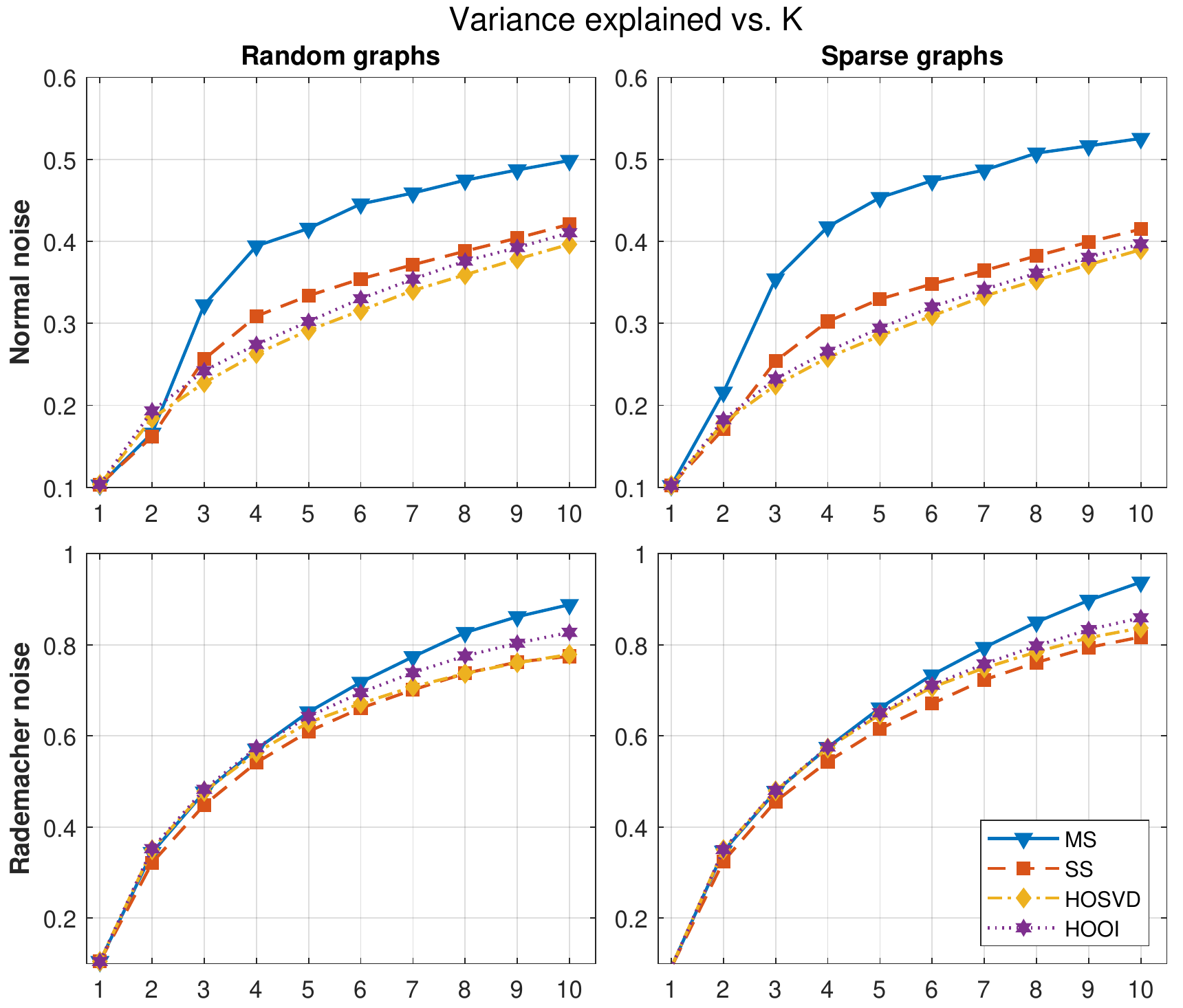}
\caption{Estimation of latent factors under various conditions. Columns specify graph structure and rows specify noise structure.}
\label{fig:var_exp}
\end{figure}

\section{Results}

Each subsection of the results highlights a different advantage of multi-scale modelling over single-scale alternatives. First, simulations are used to demonstrate improved estimation of latent factors in the presence of serious noise. Following our simulations is an analysis of $118$ healthy individuals from the HCP whom were all processed during the early stages of \cite{cole2020surface}. A method for detecting group connectome differences with varying trait values is showcased; in particular, we use multi-scale modelling to detect a relationship between reduced connectivity and increased binge drinking which appears to be unavailable to single-scale methods (\autoref{fig:mech}). We close by demonstrating improved trait predictions and greater sensitivity for detecting connectome differences.

\subsection{Simulations}

We conduct a simulation study demonstrating how multi-scale modelling yields more accurate estimation of latent factors in the presence of noise. The majority of noise in connectomics applications occurs during imaging. Here one encounters radio frequency (RF) emissions due to thermal motion inside the brain, general measurement error, and significant head movements [\cite{noise1}; \cite{noise2}].

Our simulations span two types of graph structures (random and sparse) and two types of noise (random and Rademacher). To begin, adjacency matrices were simulated at scales $\{25, 50, 75\}$, each representing $N=100$ subjects with a true rank of $10$. The common latent factors were drawn entrywise from a $N(0,1)$ distribution and the network components were drawn entrywise from a Gamma($1,1$) distribution. Sparse tensors were created by thesholding 75\% of the network modes. All matrices were normalized at this stage, and then noise was added. In half the experiments this was centered normal noise with the standard deviation taken to be one third of the signal range; in the other half of the experiments it was Rademacher noise created by randomly flipping 25\% of the graph edges for each subject.

Our metric for the performance of an estimate $\widehat{\mathbf{U}}$ of $\mathbf{U}$ is the variance explained, which we define to be $||P_{\widehat{\mathbf{U}}}\mathbf{U}||_2/||\mathbf{U}||_2$. If $\widehat{\mathbf{U}}$ is an excellent reconstruction of $\mathbf{U}$ then $P_{\widehat{\mathbf{U}}}\mathbf{U}\approx\mathbf{U}$, so the variance explained should be close to $1$. For each $K=1,\ldots,10$ we recorded the multi-scale variance explained along with the mean variances explained for TN-PCA, HOSVD, and HOOI. This test was repeated $10$ times; the average results are reported in \autoref{fig:var_exp}.

Multi-scale modelling results in favorable performance in all of these cases, with the largest gains visible under severe normal noise. This makes intuitive sense: the more data we have, the more likely we are to be able to differentiate signal from noise. Joint modelling across scales simply leverages more data for each estimate of the latent factors. We also point out that common single-scale methods suffer from identifiability issues in estimating latent factors - that is, repeated runs can produce ${\bf U}$ estimates with an arbitrary sign, making it difficult to combine estimates across scales. This issue is avoided with multi-scale modelling.

\subsection{Discovering Group Connectome Differences}

One of our primary goals is to improve methodology for relating brain structure to other variables measured on the individual, ranging from neuropsychiatric conditions and cognitive traits to exposures.  We would like to understand differences in the brain structure of individuals having low and high values of a trait of interest, while allowing for variability in brain structure across individuals in each group.

\begin{figure}[p!]
    \subcaptionbox{Single-scale (1,1) connectome changes.\label{fig:ss11}}[0.45\textwidth]{\includegraphics[width=0.45\textwidth]{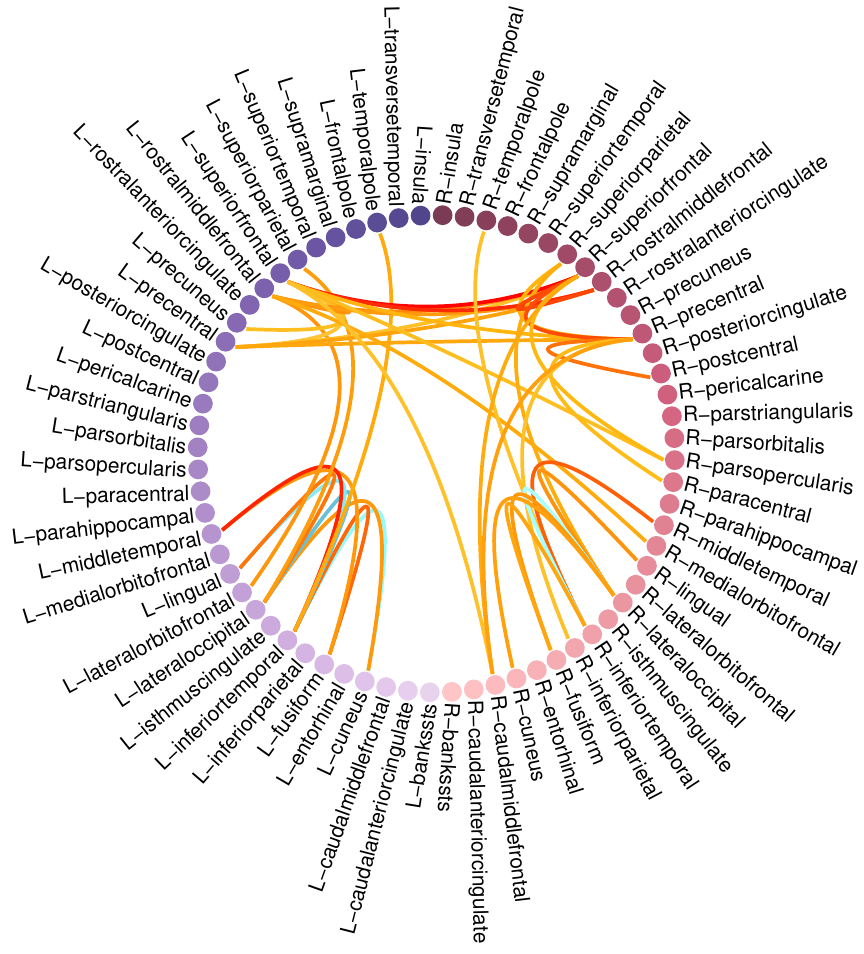}}
    \hskip -1ex
    \subcaptionbox{Multi-scale (1,1) connectome changes.\label{fig:ms11}}[0.45\textwidth]{\vspace*{3em}\includegraphics[width=0.45\textwidth]{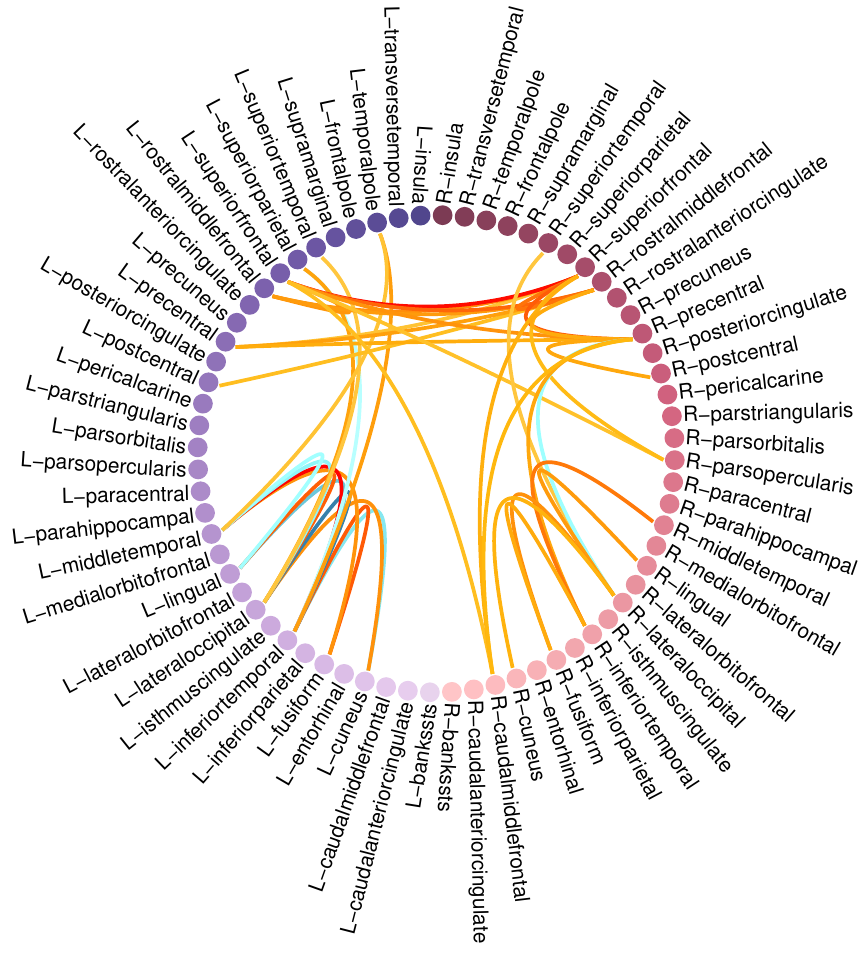}}    
    \hskip -3.6ex
    \subcaptionbox*{}[0.1\textwidth]{\vspace*{3em}\includegraphics[width=0.07\textwidth]{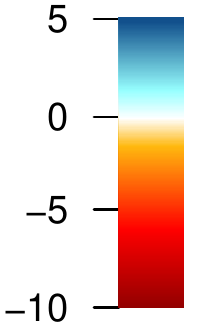}}
    \vskip -5ex
    \subcaptionbox{Single-scale (2,4) connectome changes.\label{fig:ss24}}[0.45\textwidth]{\includegraphics[width=0.45\textwidth]{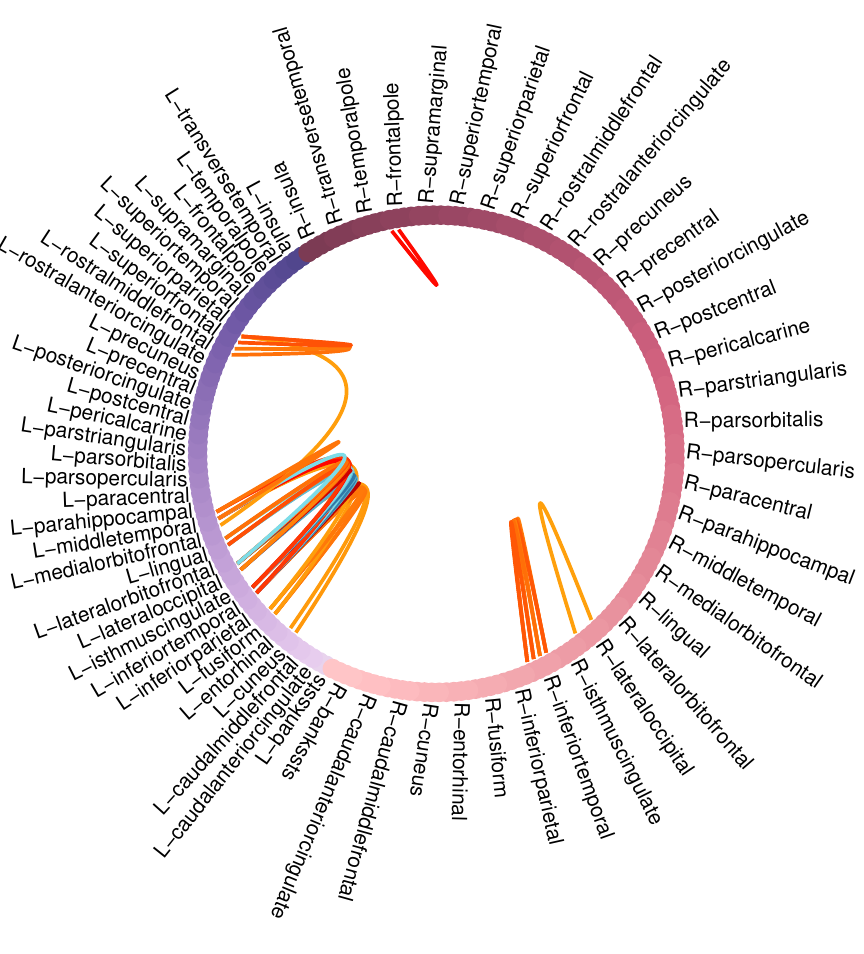}}
    \hskip -1ex
    \subcaptionbox{Multi-scale (2,4) connectome changes.\label{fig:ms24}}[0.45\textwidth]{\vspace*{3em}\includegraphics[width=0.45\textwidth]{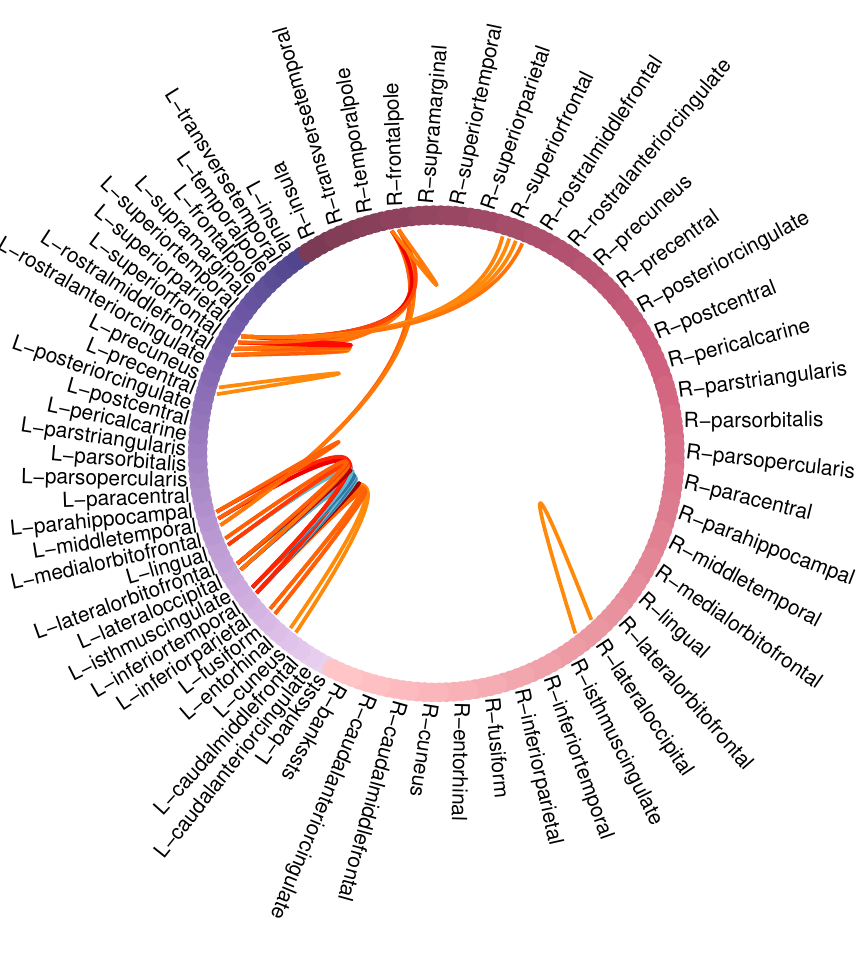}}    
    \hskip -3.6ex
    \subcaptionbox*{}[0.1\textwidth]{\vspace*{3em}\includegraphics[width=0.07\textwidth]{fig/leg.pdf}}
    \caption[.]{Largest changes in connectivity with increased lifetime binge drinking. Purple nodes correspond to the left hemisphere; red nodes correspond to the right hemisphere. Networks were multiplied by a scale-dependent constant for a more readable legend.}
    \label{fig:mech}
\end{figure}

In this subsection we showcase a method tailored for such applications and then present an example with a specific trait: worst lifetime binge drinking. We use a symmetric and an asymmetric parcellation; combining these two distinct views to the brain leads to discovery of novel edge changes across groups not readily apparent from single-scale methods. Our symmetric parcellation of choice is the Desikan atlas - $(1,1)$ in our notation - and our asymmetric parcellation is $(2,4)$, corresponding to splitting every region in the left hemisphere into two equally sized subregions and every region in the right hemisphere into four equally sized subregions.

Inferences on how connectomes change with traits is performed by finding a direction ${\bf w}\in \mathbb{R}^K$ in the embedding space that is highly correlated with a trait, and then mapping that direction back on to the space of brain networks. If ${\bf Y}$ denotes the centered trait scores and ${\bf U}$ denotes the centered latent factors, then this is equivalent to choosing 
\begin{align*}
    {\bf w} &= \underset{||{\bf \widetilde{w}}||_2=1}{\text{argmax  }}{\bf \widetilde{w}^TU^TY}.
\end{align*}

When the trait is continuous this may be easily done with canonical correlation analysis (CCA) [\cite{cca}]; when the trait is categorical we instead use linear discriminant analysis (LDA) [\cite{lda}]. The vector ${\bf w}$ captures how connectomes change with the trait score - we can then map this back onto the brain network via
\begin{align*}
    {\bf \triangle}^{(j)} &= s\sum_{h=1}^K d_h^{(j)} {\bf v}_h^{(j)}\circ {\bf v}_h^{(j)} {\bf w}_{h},
\end{align*}
where $s>0$ is a scaling parameter. Defining $s = {\bf w^TU^TY}/||{\bf Uw}||_2^2||{\bf Y}||_2^2$ to be the correlation between the projected scores and the traits allows one to compare networks across traits. Note that by using different ${\bf v}^{(j)}$ we inspect the connectome change ${\bf \triangle}^{(j)}$ at different scales, however we advocate always presenting network plots on the finest scale possible so as not to inadvertently censor potentially insightful trait-connectome relationships.

\autoref{fig:mech} illustrates this process for the ordinal ``worst binge drinking'' trait, which takes values from 1 to 7 for males and 1 to 6 for females. A value of 1 indicates no more than three drinks in a day; 2 indicates four to six drinks; 3 indicates seven to nine; 4 indicates ten to twelve; 5 indicates thirteen to fifteen; 6 indicates sixteen to twenty for males and over sixteen for females, and 7 indicates over twenty-one drinks in a single day for males. We treat this trait as continuous so that we may use CCA. Single- and multi-scale latent factors were computed using $K=10$; the resulting $\boldsymbol{\triangle}$ networks were thresholded to retain the $100$ entries with the largest absolute values at different scales. Column (a) and (c) shows the results using the single-scale TN-PCA method [\cite{tn-pca}] and column (b) and (d) shows the results from our method.

At the coarse scale, we see similar patterns in the results of our method and the single-scaled TN-PCA method: the $\boldsymbol{\triangle}$ network within each hemisphere contains mixed positive and negative values; between-hemisphere connections are dominated by negative values, focusing on the frontal cortex band. The results indicate that increased alcohol intake is related to decreased connection strength between most of brain regions, which is consistent with the literature [\cite{tn-pca}, \cite{Moselhy2001}]. However, we see multi-scale modelling reveals additional negative connectivity changes between the left rostral middle frontal gyrus, superior frontal gyri and the right frontal and right temporal lobes, which are missed by the single-scale alternatives. Importantly, at the higher resolution, our multi-scale $(2,4)$ plot in panel (d) presents a more detailed picture of the information contained in the connection between ROI pairs compared with the signal scale method. Careful inspection of the connection reduction between the left and right superior frontal ROIs indicates that the reduction in connectivity specifically happens between one half of the left superior frontal gyrus and all four subregions of the right superior frontal gyrus. These four new edges all have different weights, indicating that the connectivity changes are most relevant to specific subregions of the right superior frontal gyrus. One could continue to add finer scales to our multi-scale model to shine more light on this finding.  

\begin{figure}[t!]
    \subcaptionbox{Histograms of MSE improvements. The left plot has one omitted outlier around $-200$. \label{fig:mse_dists}}[1\textwidth]{\includegraphics[width=0.87\textwidth]{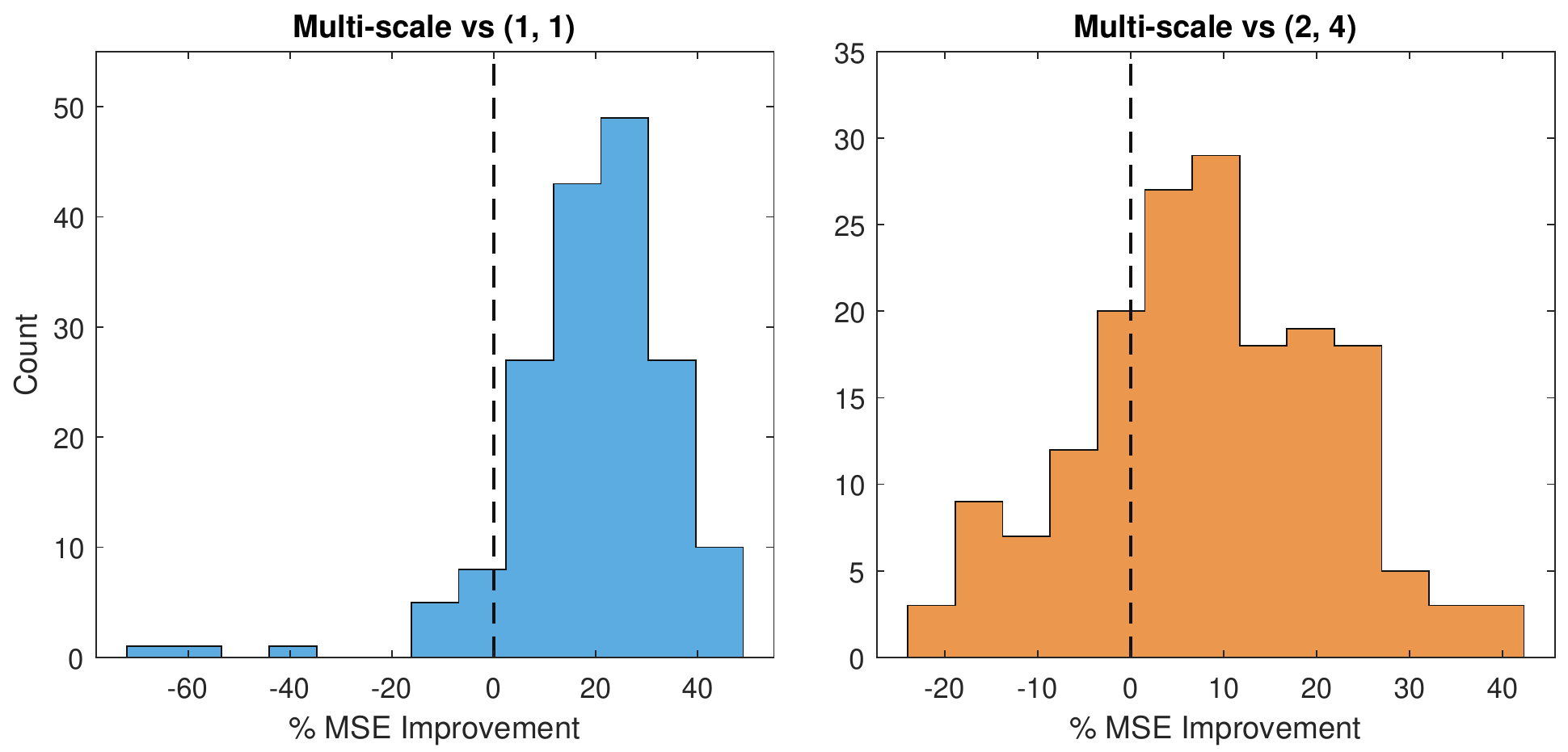}}
    \vskip 1ex
    \subcaptionbox{Highlighting improvements in predicting traits specifically related to alcohol use. \label{fig:mse_alc}}[1\textwidth]{\includegraphics[width=0.9\textwidth]{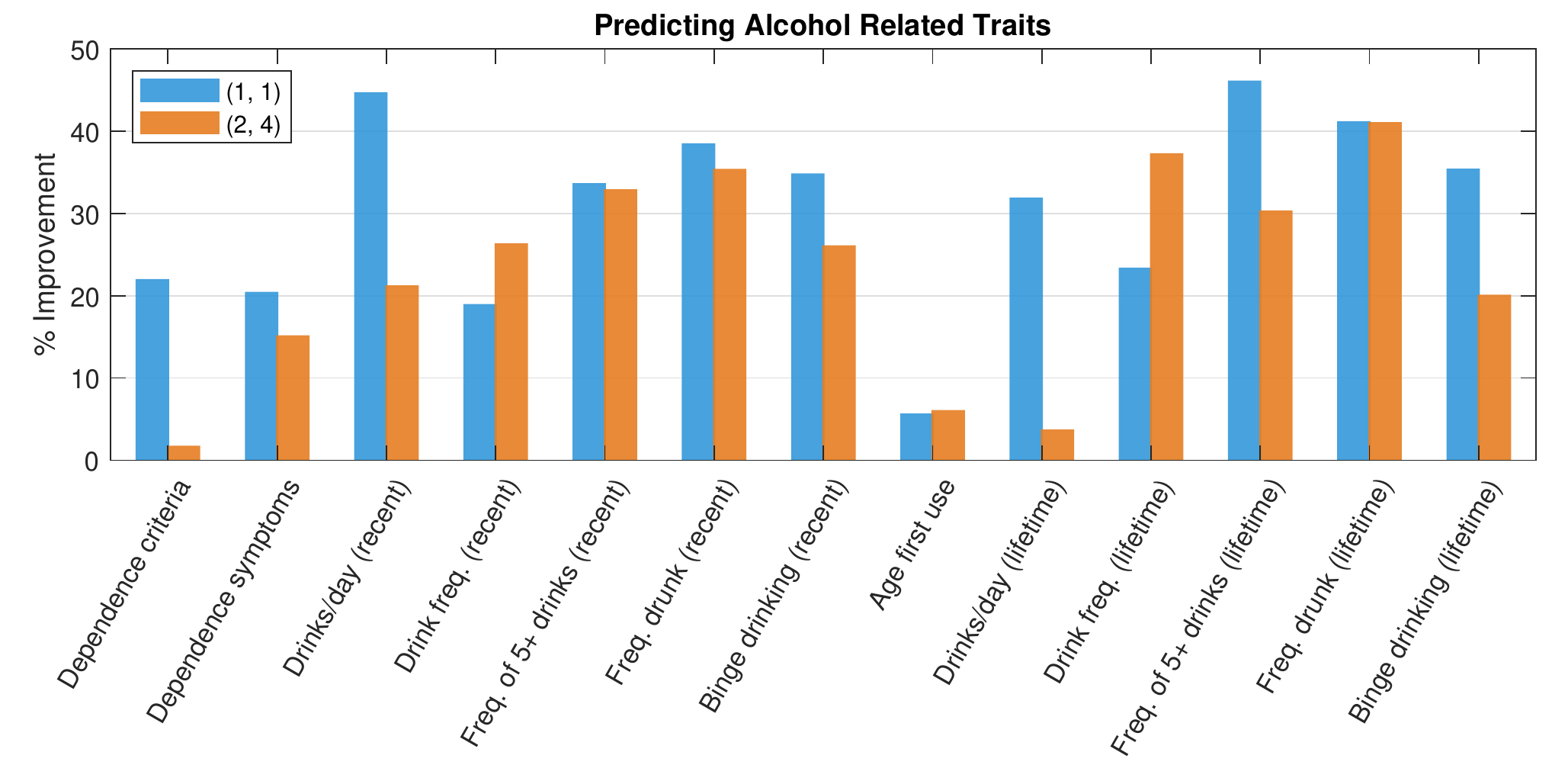}}
    \caption[.]{Visualizing the improved predictions from multi-scale methods. All calculations use the median MSE.}
    \label{fig:mse}
\end{figure}

\subsection{Improved Trait Predictions}
The synthesis of both symmetric and asymmetric graph data  yields significant improvements in predictive performance over comparably interpretable single-scale methods. We highlight this using the multi-scale $\{(1,1), (2,4)\}$ data from 118 random HCP individuals. These individuals were randomly split into a 70\% training set and a 30\% test set; a ridge regression model was then trained for each trait using the first $70$ latent factors. Predictions were made on the test set, and the mean squared error (MSE) was recorded. Overall, we repeated this test $100$ times with different splits. \autoref{fig:mse} shows distributions of the relative changes in median MSE, as well as highlights a few specific traits related to alcohol use. Our multi-scale model achieved a lower median MSE for 88.8\% of traits when compared to the single-scale $(1,1)$ model and 73.1\% when compared to the single-scale $(2,4)$ model. 

The negative changes in MSE are largely confined to traits within the domains of education, income, dexterity/endurance, delay discounting, and negative affectivity. The positive changes include the majority of traits related to substance use and dependence (alcohol, tobacco, marijuana, and hard drugs), psychiatric function (attention, aggression, anxiety, depression, insomnia, etc), sensory ability (vision, touch, taste, hearing, smell), emotion recognition, measures of healthy relationships, and family medical history.

\begin{figure}[t!]
\hspace*{0.6cm} \includegraphics[width=0.8\linewidth]{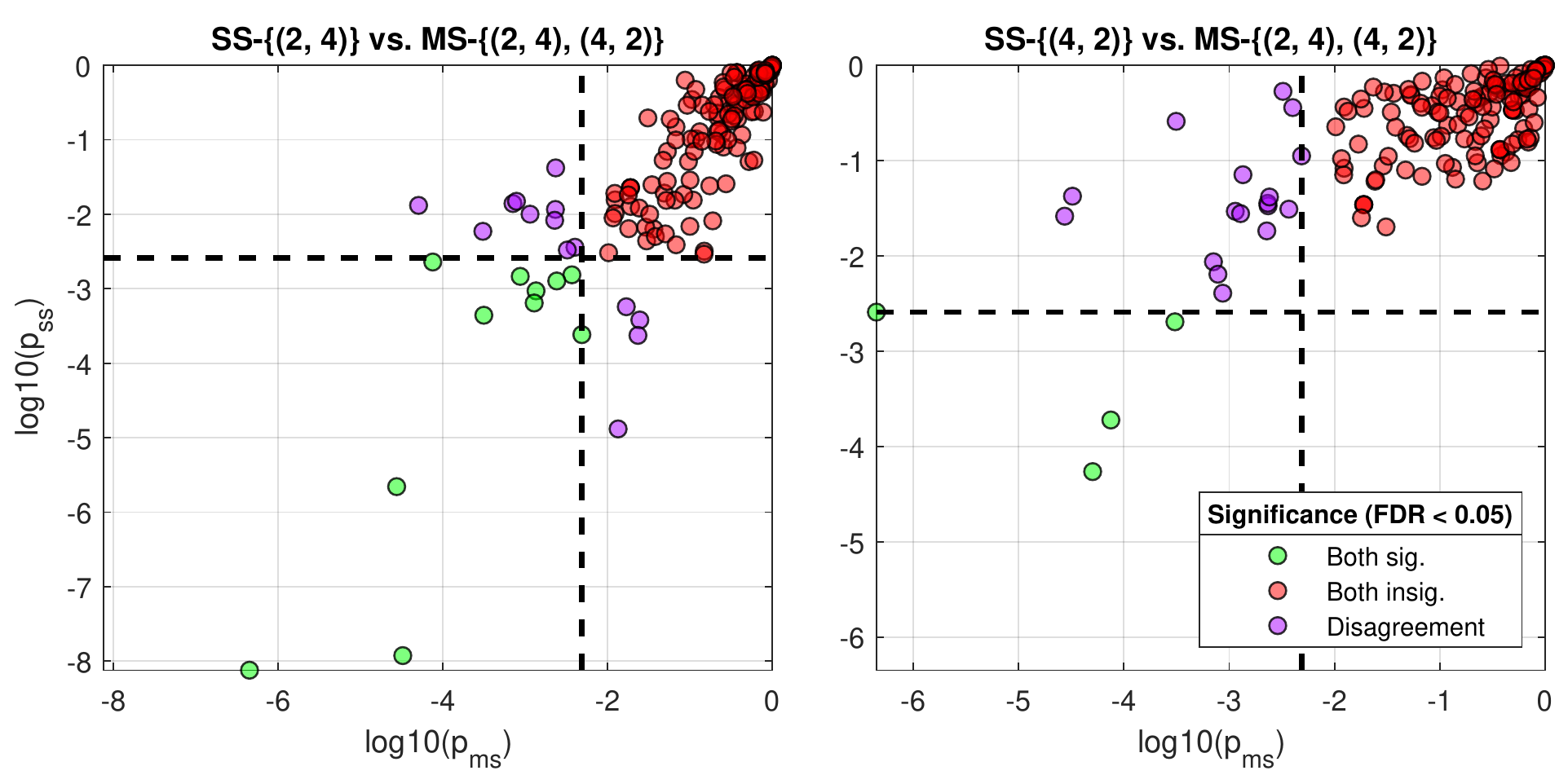}
\caption{Visualizing $p$-values from comparing $K=20$ latent factors with the MMD test. Horizontal and vertical lines correspond to FDR $<0.05$ for single- and multi-scale $p$-values respectively.}
\label{fig:mmd}
\end{figure}

\subsection{Connectome Hypothesis Testing via Latent Factors}

We can indirectly measure the significance of connectome-trait relationships by performing a hypothesis test for differences in the distribution of latent factors between individuals with high trait values and low trait values. Single-scale models are at a significant theoretical disadvantage to multi-scale models in this context for two main reasons. Firstly, they greatly amplify the number of comparisons being made. If we have $R$ parcellations and we analyze each trait separately for each parcellation, then we are performing $175\times R$ hypothesis tests; a multi-scale model leveraging all $R$ parcellations only requires $175$ tests. Secondly, $p$-values from single-scale methods cannot be interpreted on the connectome level. Hypothesis tests performed with a single discretization of the cortical surface can only be used for inference related to that discretization; multi-scale modelling frees one from these confines and allows for more general, connectome-level tests. 

We used the maximum mean discrepancy (MMD) test to detect differences in the latent factors between individuals with trait values in the upper quartile $25\%$ and those in the lower quartile (groups of $N\in [25,30]$ for our $118$ unique individuals based on data availability and tied values). \autoref{fig:mmd} visualizes $p$-values for the multi-scale $\{(2,4), (4,2)\}$ model against a single-scale alternative, with horizontal and vertical lines corresponding to significance thresholds after adjusting for a false discovery rate (FDR) [\cite{fdr}] of no more than $5\%$. 

In particular when $K=20$ our multi-scale model detects significantly different latent factor distributions for two traits related to fluid intelligence (Penn matrix test overall score and number of skipped questions), two traits related to reading ability (oral reading test scores, both age adjusted and unadjusted), two traits related to working memory (list sorting test scores, both age adjusted and unadjusted), and two traits related to overall psychiatric function (normalized anxiety/depression scores and raw DSM hyperactivity scores). None of these traits were significant under the associated single-scale models.

\section{Discussion}

In this paper we develop a new method for summarizing an individual's structural connectome using multiple discretizations of the cortical surface. Our proposed approach is tailored for interpretable analysis of relationships between brain structure and individual specific factors. By using data from multiple parcellations simultaneously, we can potentially improve power to detect relationships between connectomes and traits; for example, we observed a significant association between the connectome and binge drinking that was not apparent in single scale analyses.  In addition, we demonstrated significantly improved predictive performance over comparable single-scale models, as well as generally more accurate detection of connectome differences and greater robustness to different types of noise. All code is available, ready-to-use, on GitHub.

Our algorithm performed best when we combined both low and high-resolution data. This makes intuitive sense: high-resolution data contain a great deal of information, but analysis is generally complicated by problems such as individual ROI misalignment and sparsity. Low-resolution data contain less information, but individuals are extremely well aligned and the observed networks are not sparse. Anchoring high-resolution data to low-resolution data via common latent factors allows us to leverage information only available at fine scales without being overwhelmed by the aforementioned complications.

Another consistent theme in our experiments was disappointing results with symmetric parcellations; surprisingly, asymmetric parcellations almost always yield better performance - even when comparing among single-scale models. We believe this indicates that future research on defining optimal brain parcellations is paramount. Perhaps ROIs can be considered as network communities which can be learned from the ultra-high resolution adjacency matrix representation of the connectome. Different parcellations could be found by running e.g. hierarchical community discovery [\cite{reichardt2006statistical}; \cite{lancichinetti2009detecting}] with different tuning parameters, similarity metrics, and loss functions. One could imagine creating similarity metrics or loss functions that factor in trait information; for example detecting a set of communities with high interconnectivity for individuals with the highest trait values. Such methods could produce a wide variety of ``optimal'' parcellations.

The above approach may generate many parcellations that are optimal in some sense, but it is unlikely that all of these parcellations are necessary for multi-scale inference. It is therefore also important to develop methods for selecting ideal subsets of multi-scale data. Hypothesis testing provides a simple solution: the MMD test can be applied to see if the latent factors change significantly after adding a new scale. Another relatively simple solution may involve adding weights to each term of the objective function in our model; these weights could then be estimated in some supervised way, with thresholding used to eliminate irrelevant parcellations. Finally, it may be possible to define an influence function in terms of the latent factors, which would provide a clear interpretation of the effect of new parcellations as well as opening the door to analyzing the effects of head movement on inference.

\section*{Acknowledgments}
This research is partially supported by grant 1R01MH118927-01 of the United
States National Institute of Health (NIH). Data collection and sharing for this project was provided by the MGH-USC Human Connectome Project (HCP; Principal Investigators: Bruce Rosen, M.D., Ph.D., Arthur W. Toga, Ph.D., Van J. Weeden, MD). HCP funding was provided by the National Institute of Dental and Craniofacial Research (NIDCR), the National Institute of Mental Health (NIMH), and the National Institute of Neurological Disorders and Stroke (NINDS). HCP data are disseminated by the Laboratory of Neuro Imaging at the University of Southern California.

\newpage 

\bibliography{ref,paperdti.bib}

\newpage 

\appendix
\section{Imputing missing data}

The multi-scale graph model can naturally impute missing data for a subset of the subjects at any scale. Adjusting the algorithm to accommodate this requires more notation, which we clarify with a simple example. Consider an experiment with $N=10$ subjects where brain imaging data are collected at finer scales each day for three days. Assume that subject $10$ is unable to come in on the third day. This presents no issue: we can compute the subject modes of individuals $1-9$ using their data across all three days. Individual $10$'s subject mode can be computed using the data of individuals $1-10$ over the first two days, and network modes can be computed using complete subsets of data within each scale. 

Formally, assume we have tensors $\mathcal{X}^{(j)}\in \mathbb{R}^{P_j\times P_j \times N}$ with $j=1,...,R$. Define $\mathbf{\Gamma}\in \mathbb{R}^{N\times R}$ by the rule 
\begin{align*}
    \mathbf{\Gamma}_{ij} = \begin{cases}
1 & \text{if we have data for subject }i \text{ at scale }P_j \\
0 & \text{else}
\end{cases}.
\end{align*}
For each subject $\ell=1,...,N$, define
\begin{align*}
    S(\ell) &= \{w : \mathbf{\Gamma}_{\ell w} = 1, w=1,...,R\}
\end{align*}
which is the set of indices for the tensors that contain information about subject $\ell$. In the above example, $S(9)=\{1,2,3\}$ and $S(10)=\{1,2\}$. Lastly for $\ell=1,...,N$, set
\begin{align*}
    G(\ell) = \{z:S(\ell)\subseteq S(z), z=1,..,N\}.
\end{align*}

This is the set of individuals with at least as much information as individual $\ell$. In the above example we have $G(1)=\cdots =G(9)=\{1,...,9\}$ and $G(10) = \{1,...,10\}$. If $G(\ell)$ is nonempty then we are free to update individual $\ell$'s latent factor as follows:
\begin{align*}
(\widehat{\mathbf{u}}_h)_\ell|\mathbf{v}_h^{(1)}, ..., \mathbf{v}_h^{(R)} &= \bigg(E_{\max}\bigg(\sum_{i=1}^R\bigg(\mathcal{X}_{::G(\ell)}^{(i)} \times_1 \mathbf{P}_{h-1}^{(i)}\mathbf{v}_h^{(i)} \times_2 \mathbf{P}_{h-1}^{(i)}\mathbf{v}_h^{(i)}\bigg)\\
&\hphantom{--------}\bigg(\mathcal{X}_{::G(\ell)}^{(i)} \times_1 \mathbf{P}_{h-1}^{(i)}\mathbf{v}_h^{(i)} \times_2 \mathbf{P}_{h-1}^{(i)}\mathbf{v}_h^{(i)}\bigg)^T\bigg)\bigg)_\ell
\end{align*}

This update can be done in blocks: one only needs to compute a total of $\#\{S(\ell) : \ell=1,...,N\}$ different matrices and find the maximum eigenvalue of these. Furthermore, rank one SVD updates may be used to compute the subject modes for groups of subjects that differ by only one missing scale, resulting in only minor increases in computational complexity [\cite{stange}].

The network modes are updated using complete subsets data at each scale:
\begin{align*}
        \widehat{\mathbf{v}}_h^{(j)}|\mathbf{u}_h &= E_{\max}\bigg(\mathbf{P}_{h-1}^{(j)}(\mathcal{X}_{::\{z:\Gamma_{zj}=1\}}^{(j)} \times_3 (\mathbf{u}_h)_{\{z:\Gamma_{zj}=1\}})\mathbf{P}_{h-1}^{(j)}\bigg),
\end{align*}
which comes with no increase in computational complexity. This generalization will produce Kruskal decompositions $\{\mathbf{d}^{(j)},\mathbf{V}^{(j)}, \mathbf{U}\}$ as in the complete data case, allowing for automatic imputation of missing data by reconstructing the $\mathcal{X}^{(j)}$.

\end{document}